\documentclass[twocolumn,showpacs,showkeys,showkeywords,preprintnumbers,amsmath,amssymb]{revtex4}

\usepackage{graphicx}
\usepackage{dcolumn}
\usepackage{bm}

\newcommand{\newc}{\newcommand}
\newc{\ie}{{\it i.e.}}
\newc{\eg}{{\it e.g.}}

\def\MT2{M_{T2}}

\def\m0{${0}$}

\def\tg{{\tilde g}}

\newcommand{\neut}{$\chi$}

\def\MT2{M_{T2}}

\def\m0{${0}$}

\def\tg{{\tilde g}}

\def\NPB#1#2#3{{\rm Nucl. Phys.} B {\bf{#1}} (#2) #3}

\def\PR#1#2#3{{\rm Phys. Rep.} {\bf{#1}} (#2) #3}
\def\PRD#1#2#3{{\rm Phys. Rev.} D {\bf{#1}} (#2) #3}

\def\PRL#1#2#3{{\rm Phys.~Rev.~Lett.} {\bf{#1}} (#2) #3}
\def\APP#1#2#3{{\rm Astropart.~Phys.} {\bf{#1}} (#2) #3}
\def\APJ#1#2#3{{\rm Astrophys.~J.} {\bf{#1}} (#2) #3}

\def\AA#1#2#3{{\rm Astronomy and Astrophysics} {\bf{#1}} (#2) #3}
\def\NATURE#1#2#3{{\rm Nature} {\bf{#1}} (#2) #3}
\def\MNRAS#1#2#3{{\rm Mon. Not. Roy. Astron. Soc.} {\bf{#1}} (#2) #3}
\begin{document}
\preprint{APS/123-QED}
\title{Complementarity of Gamma-ray and LHC Searches for Neutralino Dark
Matter \\in the Focus Point Region}

\author{E.~Moulin$^1$}
\email{emmanuel.moulin@cea.fr} \altaffiliation[Also at
]{IRFU(ex-DAPNIA)/SPP, CEA Saclay, F-91191 Gif-sur-Yvette, Cedex,
France}
\author{A.~Jacholkowska$^1$}
\author{G.~Moultaka$^1$}
\author{J.-L.~Kneur$^1$}
\author{E.~Nuss$^1$}
\author{T.~Lari$^2$}
\author{G.~Polesello$^3$}
\author{D.~Tovey$^4$}
\author{M.~White$^5$}
\author{Z.~Yang$^6$}
\affiliation{$^1$ Laboratoire de Physique Th\'eorique et
Astroparticules, CNRS/IN2P3 et Universit\'e Montpellier 2, Place
Eug\`ene Bataillon, F-34095 Montpellier Cedex, France}
\affiliation{$^2$ INFN, Sezione di Milano, Via Celoria 16, I-20133
Milano, Italy} \affiliation{$^3$ INFN, Sezione di Pavia, Via Bassi
6, I-27100 Pavia, Italy} \affiliation{$^4$ Department of Physics
and Astronomy, University of Sheffield, Hounsfield Road, Sheffield
S3 7RH, UK} \affiliation{$^5$ Cavendish Laboratory, University of
Cambridge, Madingley Road, Cambridge, CB3 0HE,
UK}\affiliation{$^6$ Department of Physics, Carleton University,
Ottawa, ON K1S 5B6, Canada}

\date{\today}
\begin{abstract}
We study the complementarity between the indirect detection of
dark matter with $\gamma$-rays in H.E.S.S. and the supersymmetry
searches with ATLAS at the Large Hadron Collider in the Focus
Point region within the mSUGRA framework. The sensitivity of the
central telescope of the H.E.S.S.~II experiment with an energy
threshold of $\rm \sim$ 20 GeV is investigated. We show that the
detection of $\gamma$-ray fluxes of $\rm
\mathcal{O}(10^{-12})\,cm^{-2}s^{-1}$ with H.E.S.S. II covers a
substantial part of the Focus Point region which may be more
difficult for LHC experiments. Despite the presence of multi-TeV
scalars, we show that LHC will be sensitive to a complementary
part of this region through three body NLSP leptonic decays. This
interesting complementarity between H.E.S.S. II and LHC searches
is further highlighted in terms of the gluino mass and the two
lightest neutralino mass difference.
\end{abstract}
\pacs{95.35.+d, 11.30.Pb, 12.60.Jv, 95.55.Ka}
\keywords{Dark Matter, Supersymmetry, SUGRA, Cherenkov telescopes,
LHC physics, relic density}
\maketitle

\section{Introduction}
\label{sec:intro} A preferred particle physics candidate for the
Dark Matter (DM) component in the Universe is an electrically
neutral Lightest Supersymmetric Particle (LSP), which in various
supersymmetry (SUSY) breaking scenarii is the lightest neutralino
\neut~\cite{goldberg,jungman}. The self-annihilations of
neutralinos in halos of galaxies would produce Standard Model
particles: $\gamma$, $\nu$, charged leptons and hadrons. The
ongoing and forthcoming ground-based (H.E.S.S.~\cite{hess},
MAGIC~\cite{magic}, VERITAS~\cite{veritas} and space
(AMS~\cite{ams0}, GLAST~\cite{glast}, PAMELA~\cite{pamela})
experiments detecting galactic cosmic rays will open large new
windows for the DM annihilation signal detection. In particular,
phase II of the H.E.S.S. experiment will provide an improved
sensitivity due to a combination of a large effective area, low
energy threshold, and an excellent angular resolution
necessary for the efficient hadron background suppression.\\
Simultaneously to the new generation of the cosmic ray detectors
which will provide outstanding data in the dark matter
detection domain, the Large Hadron Collider (LHC) start-up in 2008
will explore physics in the TeV energy range and thus probe the
origin of electroweak symmetry breaking. It will also introduce a new era of
physics and particle searches beyond the Standard Model,
and in particular ought to discover the new particles predicted by
supersymmetry or other new states such as those foreseen by models with extra
spatial dimensions.  Under the
hypothesis that dark matter is composed of a SUSY (or
Kaluza-Klein) particle, a relationship between the astrophysical
signal observation and the properties of the new particles found
at LHC can be established. The information from LHC measurements
will also allow us to discard various hypotheses on the nature of
the DM~\cite{weiglein}. On the other hand, in case of a signal
detection which is not correlated with an astrophysical source,
H.E.S.S. would be able to investigate the DM distribution profile
in halos of galaxies independently of the particle physics
uncertainties. Accurate reconstruction of the morphology of the
sources could help to discard specific halo models.\\
The aim of this paper is to study some experimental and
phenomenological aspects of the complementarity between searches
of SUSY particles in the ATLAS experiment and the detection
potential of the H.E.S.S. telescope in phase II, in the case of
the so-called Focus Point (FP) region in the mSUGRA framework of
the Minimal Supersymmetric Model (MSSM). The most realistic Monte
Carlo simulations for both experiments will provide input to this
study. As the measured variables in ATLAS are related to various
mass differences of the SUSY states, the H.E.S.S. II sensitivity
will be investigated with respect to these variables. In
particular, we will investigate the dependence of the gamma-ray
fluxes and the annihilation cross-section on the gluino mass and
the mass difference of the two lightest neutralinos. Detailed
studies in the literature have investigated the complementarity of
accelerator measurements at LHC/ILC and astrophysical
observations, e.g. ~\cite{Baltz:2006fm,Hooper:2006wv}, to unveil
the nature of Dark Matter. Here, we focus on the complementary
constraints in the FP region with the forthcoming atmospheric
Cherenkov telescope H.E.S.S. II and LHC searches using realistic
performances for both experiments. This FP region is also of great
interest for searches via neutrino
telescopes~\cite{Barger:2007xf}.

The rest of the paper is organized as follows: some key points of
the supersymmetric focus point framework and related
phenomenological features are given in section 2. Section 3
describes various issues related to $\gamma$-ray signals expected
in Cherenkov telescopes and the detection sensitivity of the
upcoming phase II of the H.E.S.S. experiment. The constraints from
the $\gamma$-ray flux senstivity to the parameter space in the
focus point region are studied in section 4.  MSSM spectrum
measurements at the LHC for this parameter space region are
discussed in section 5. Section 6 is devoted to the potential
complementarity between H.E.S.S. II and ATLAS measurements.
Conclusions and perspectives are given in section 7.

\section{Supersymmetric framework and the Focus point region}
\label{sec:susy}

We consider the framework where supersymmetry breaking in a hidden
sector is mediated to the visible world through gravitational
interactions between the very heavy  hidden fields and the MSSM
sector. Although little is known about the actual dynamics
responsible for this breaking and its mediation mechanisms, it is
reasonable to assume that the gross features of the low energy
spectrum can be encapsulated in a universality assumption (and
possible deviations from it) for the effective soft supersymmetry
breaking parameters at a scale close to the Planck or the GUT
scale. The more detailed features of this spectrum are then
determined by  quantum effects obtained through the running down
to the electroweak scale. It is important to keep in mind that the
theoretical uncertainties at the high (SUSY breaking) scale are
different in nature from those involving the running to the low
(electroweak) scale. The former cannot be reduced without a better
knowledge of the underlying new physics, while the latter are in
principle reducible through the improved calculational theoretical
tools such as the Renormalization Group Evolution (RGE) codes and
spectrum calculators,  where for instance scale dependencies,
threshold effects, higher order corrections, etc... are
standardized. In practice, the effects of these two types of
uncertainties are not easy to disentangle when one tries to
determine in a model-independent bottom-up approach the
fundamental parameters of the model, ultimately starting from
experimental data. Moreover, in a collider environment such as the
LHC one would presumably aim first at typical SUSY discovery
events, then at the determination of mass differences of various
SUSY particles which are kinematically accessible. With this
respect, resorting to detailed predictions in a top-down approach
should be taken only as a guide for the possible patterns of SUSY
mass spectra, production rates, branching ratios, etc..., as well
as the complementary requirements for a good dark matter candidate
(relic density, predictions for direct and indirect dark matter
searches).

In this paper we focus on MSSM spectrum patterns where all the
scalar quarks and leptons are in the multi-TeV range and thus out
of direct reach at the LHC, while the lighter neutralinos and
charginos and of course the lightest Higgs remain accessible. Such
patterns are motivated by the so-called focus point
scenarios~\cite{FMM} where it is noted within mSUGRA that for
$\tan \beta \gtrsim 5$ and $m_{1/2}, A_0 \lesssim 1$ TeV, $m_0$
could be very large and still allow electroweak symmetry breaking
and the right $Z$ boson mass to occur without a large fine-tuning
between the order parameter $m_{H_u}^2 (\le 0)$ and $\mu^2 (\ge
0)$ at the electroweak scale. This feature is due to the quantum
effects which push the running of $m_{H_u}^2$ to an essentially
unique value, a focus point, at some given low scale, irrespective
of its initial value at very high scales. It so happens that for
the above range of parameters this focus point occurs at a scale
$Q_0$ of order the electroweak scale, and where $|m_{H_u}(Q_0)|$
is also of the same order, together with $\mu^2(Q_0) (\sim
m_{H_u}^2(Q_0))$ due to electroweak symmetry breaking (EWSB). In
this case, the smallness of the $\mu$ parameter leading to a non
negligible higgsino component for the lightest neutralino has
immediate consequences on the dark matter issues~\cite{FMW,BKPU},
reducing the neutralino relic density and increasing the reaction
rates and fluxes respectively for direct and indirect dark matter
detection. These distinctive features, together with other ones
due to the heaviness of the sfermion sector (such as the
suppression of  flavor changing neutral currents and the electric
dipole moments of the  electron and the neutron) makes this
scenario with low fine-tuning very interesting to study. However,
the low fine-tuning in the EWSB condition seems to be traded for a
very large sensitivity to the top yukawa coupling and thus to the
physical top mass, as was shown in various studies
in~\cite{BKTplus} (see also~\cite{BKPU}). The practical
consequence of this sensitivity is the important dispersion in the
output of different spectrum calculator and RGE codes and thus in
the predictions of the various experimental observables, not to
mention numerical convergence issues as pointed out in~\cite{BKPU}
through a comparison of the two codes ISAJET 7.69~\cite{isajet}
and SUSPECT 2.34~\cite{suspect}. These features hint at the need
for further standardisation of the various codes through the
inclusion of improved calculations at similar levels; but we
anticipate that this would probably not be enough  to improve
substantially the situation, due to the intrinsic high sensitivity
on input parameters in the focus point regions. For that, one
would require a better theoretical understanding of this high
sensitivity (meaning of the dependence of the focus point location
on the running scale, etc). Since in the present study we are
mainly interested in the potential of the experimental
complementarity between H.E.S.S. phase II and ATLAS for the
challenging multi-TeV spectrum pattern, we will  ignore the above
theoretical uncertainties and consider the focus-point-like
regions sticking for convenience to a given code, namely
ISAJET~\cite{isajet} to generate the MSSM mass spectrum and
couplings. We have however performed cross-checks with the SUSPECT
code~\cite{suspect}  and found reasonable agreement for the
overall focus point region (see also section 4 for further
discussions).  The mSUGRA parameter space is defined by the
following set of four parameters and a sign\footnote{the acronym
mSUGRA is used everywhere in this  paper in a loose sense, that is
without assuming a GUT  scale relation between the soft parameters
$A_0$ and $B_0$ that would result from a flat K\"ahler metric.}:
\begin{eqnarray}
\label{eqn:msugra}
 m_0,\,m_{1/2},\,A_0,\,\tan \beta,\,sign(\mu)
\end{eqnarray}
with $m_0$ the common scalar mass, $m_{1/2}$ the common gaugino
mass, $A_0$ the trilinear coupling, $sign(\mu)$ the sign of the
higgsino mass and $\tan \beta$ the Higgs vacuum expectation values
(VEV) ratio. In the sequel of the paper, we will perform detailed
scans of the parameter space in the focus point region using both
ISAJET 7.69 interfaced with DarkSUSY 4.1~\cite{darksusy}, and
ISAJET 7.71.

\section{Indirect detection of Dark Matter}
\label{sec:indirect}
\subsection{Galactic sources of dark matter halos}
The choice  of the astrophysical targets to be observed is of main
importance for the dark matter signal searches. Among the most
commonly suggested
sources~\cite{bergstrom,baltz,fornengo,tasitsiomi0}, various
studies such as~\cite{bergstrom} have shown that the Galactic
Center could be an interesting object under the assumption that
the DM density halo presents a strong density increase towards its
inner region. The dark matter density profile extrapolated to the
central region is currently parameterized following the cored
spherical models as proposed by~\cite{evans1} or as halos with a
cusp as proposed by~\cite{nfw} and~\cite{moore}. The cuspiness of
the halo is not yet confirmed by the rotation curves of the
stars~\cite{binney1} and is in contradiction with a high value of
the microlensing optical depth in the central
region~\cite{debattista} which cannot be due to the diffuse DM but
rather to the presence of faint stars and brown dwarfs. The other
argument against the presence of a large density of the DM in the
Galactic Center is its impact on the large galactic bar rotation
which could be modified by the dynamical friction with dark matter
medium~\cite{binney2}. However, microlensing optical depth and
dynamical friction arguments are slightly
controversial~\cite{mcmillan} and a cusp at the Galactic Center is
not yet ruled out. It has to be underlined that the Galactic
Center region is rich in potential $\gamma$-ray emitters such as
supernova remnants, the supermassive black hole Sgr A*, the
recently discovered HESS J1745-290 source~\cite{aharonian1}, and
the standard interactions of charged cosmic rays producing diffuse
$\gamma$-ray emission~\cite{aharonian3}.

The most promising astrophysical objects with large
mass-to-luminosity ratio are the Sphero\"idal Dwarf Galaxies
(dSph) of the Local Group, periodically crossing the Galactic
Plane. The not too distant ones are Draco, Sagittarius and Canis
Major located respectively at $\sim$ 80, 24 and 8 kpc from the
heliocentric position. The presently measured star rotation curves
provide the mass-to-luminosity ratio in the range of 20 to 100 and
no final answer has been given about cuspiness or not of the dark
matter density profile, so the cored density profile should also
be considered when predictions for $\gamma$-ray fluxes from the
neutralino annihilations are performed. As the line-of-sight
integration of dark matter density is mainly driven by the
distance value of the object, the Canis Major galaxy is certainly
the most promising target. On the other hand, this galaxy seems to
be the most disrupted one by the tidal forces as it orbits our
Milky Way. The expected flux ratios between dSphs and the Galactic
Center taken as a reference source are respectively $\sim$ 0.03,
0.15 and 0.5 for Draco, Sagittarius and Canis Major, as calculated
for a Navarro-Frenk-White (NFW) cusped profile by~\cite{evans2}.
Despite smaller $\gamma$-ray fluxes expected for dwarf galaxies,
they consist of less complex environments compared to the Galactic
Center region as pointed above and present lower background coming
from astrophysical sources. The extended TeV emission along the
galactic plane leads to a diffuse astrophysical background which
is very challenging to overcome. Dwarf galaxies are currently in
the scope of the H.E.S.S. experiment and first results on
Sagittarius have been reported in~\cite{sgrdwarf}.

In this paper we consider the Galactic Center as a benchmark
source in order to compare our predictions with those provided by
other authors. The $\gamma$-ray fluxes from annihilation of
neutralinos in a spherical dark halo are obtained from
\begin{eqnarray}
\frac{d\Phi_{\gamma}(\Delta\Omega,E_{\gamma})}{dE_{\gamma}}\,=\,\frac{1}{4\pi}\frac{\langle\sigma
v\rangle}{2\,m_{{\chi}}^2}\,\frac{dN_{\gamma}}{dE_{\gamma}}
\times\bar{J}(\Delta\Omega)\,\Delta\Omega \label{eqn:flux}
\end{eqnarray}
with :
\begin{eqnarray}
\bar{J}(\Delta\Omega)\,=\,\frac{1}{\Delta \Omega}\int_{\Delta
\Omega}d\Omega\int_{l.o.s}\rho^2[r(s)]ds\, . \label{eqn:jbar}
\end{eqnarray}
Eq.~(\ref{eqn:flux}) is expressed as a product of a particle
physics term and an astrophysics term. The former contains
$\langle\sigma v\rangle$, the velocity-weighted annihilation
cross-section, the neutralino mass $m_{\chi}$ and the gamma-ray
annihilation spectrum $dN_{\gamma}/dE_{\gamma}$. The astrophysics
part depends on $\rho$, the radial density profile expressed in
terms of $r(s) = \sqrt{s^2+s_0^2-2ss_0\cos\theta}$ where $s$ is
the heliocentric distance of a given point along the line of sight
(l.o.s) in the Galactic halo, $s_0$ is the heliocentric distance
of the Galactic Center, and $\theta$ denotes the angle between the
direction of the Galactic Center and the l.o.s. The integral over
$s$ ranges from 0 to $s_0 \cos \theta + \sqrt{R_0^2 - s_0^2 \sin^2
\theta}$ where $R_0$ is the radial extension of the spherical
halo. The integration is performed along the l.o.s. to the target
and averaged over the solid angle $\Delta\Omega$ usually matching
the angular resolution of the detector.

\subsection{The H.E.S.S. experiment}
The High Energy Stereoscopic System (H.E.S.S.) consists of four
Imaging Atmospheric Cherenkov Telescopes (IACTs)~\cite{hess}. It
is designed to detect very high energy (VHE) $\gamma$-rays in the energy range
from 100 GeV up to 100 TeV.
H.E.S.S. detects Cherenkov light emitted from electromagnetic
cascades of secondary particles resulting from the
$\gamma$-ray/hadron primary interaction in the upper atmosphere.
Cherenkov images of air showers are used to deduce the energy and
direction of the primary particle. At the zenith, the energy
threshold of the system is 100 GeV and for point sources, an
energy resolution of 15\% is achieved with stereoscopic
measurements which determine the height of the maximum shower
development. The angular resolution for individual $\gamma$-ray is
better than 0.1$^{\circ}$ and the point source sensivity for a
5$\sigma$ detection reaches 1\% of the flux of the Crab nebula in
25 hours~\cite{crabe}. The combination of these unequalled
characteristics allows for detailed studies of high energy
$\gamma$ source morphology.

The future of H.E.S.S. is in an upgrade to the existing telescope
array. H.E.S.S. phase II~\cite{punch} consists of a very large
single telescope located at the center of the H.E.S.S. I array,
expected to start taking data as soon as end of 2008. The total
mirror collection area is about 600 m$^2$ and the camera with a
3$^{\circ}$ field of view is made up of 2048 photomultiplier
tubes. The pixel size of 0.07$^{\circ}$ will provide a better
resolved shower image as compared to H.E.S.S. I. In the
stand-alone operating mode, H.E.S.S. II will reach an energy
threshold as low as 20 GeV.

\subsection{Detection sensitivity of ground-based Cherenkov telescopes} The
phase II of the ground-based Imaging Atmospheric Cherenkov
Telescopes with its improved sensitivity at energies below 100 GeV
and excellent angular resolution, may allow exploration of
significant portions of the SUSY parameter space. Its sensitivity
will depend strongly  on the hadron background suppression
procedures that will be developed at a few tens of GeV, unlike the
space telescopes where the background is dominated by the diffuse
photon emission. At present energies of H.E.S.S. I above 100 GeV
threshold, the photon/hadron shower identification allows us to
discover weak extended sources in a reasonable time of observation
below 50 hours. One has to keep in mind that the phase II of
H.E.S.S. will provide new results in the interesting astrophysical
and/or new physics domains only if the steep hadron (proton,
nuclei, etc) and cosmic electron spectra can be discarded.
Forthcoming ACTs (such as the MAGIC II experiment \cite{magic}
with two 17 m diameter telescopes) will investigate energies below
100 GeV. However, such ACTs are not expected to lower their energy
thresholds as much as H.E.S.S. II will do, since these thresholds
scale roughly linearly with the mirrors' size.

The following observation conditions of a given source will
determine the effective sensitivity to the SUSY signal:
\begin{itemize}
\item the elevation angle as seen by H.E.S.S. on the horizon will
have strong impact on the energy threshold, the hadron background
suppression
 and energy resolution;
\item the number of active telescopes, the zenith angle
corresponding to a given source observation and its angular offset
with respect to the center of the camera, will determine the
acceptance values as a function of energy;
\item the source
spatial profile and weakness of the signal will condition the
accepted number of $\gamma$ and the power of the background
suppression.
\end{itemize}
The optimal source observation conditions are currently fulfilled
in H.E.S.S. experiment by the Galactic Center and Dwarf Spheroidal
Galaxy observations.

Typical uncertainties on the integrated flux measured with
H.E.S.S. are of the order of 20\% which account mainly for
systematics  due to different signal extraction methods. The main
contribution to these systematic effects come from the
uncertainties on the hadron background suppression factors,
strongly varying with the studied energy domain. As far as the low
energy part of the spectrum is concerned, the data to come from
H.E.S.S. phase II will allow us to explore the 20 GeV energy range
for which there is no prior experience. The values quoted here are
provided by Monte Carlo studies relying on the extrapolation from
higher energies. The impact of the energy resolution on the energy
threshold determination may be considered as negligible. The
foreseen experimental uncertainty is still much less than the
astrophysical uncertainties (see also section 4).

\subsection{Sensitivity calculation}
In the case of Cherenkov telescopes where the signal is extracted
by ON-OFF subtraction, the significance S is given by~\cite{lima}
\begin{eqnarray}
S = \frac{N_{\gamma}}{\sqrt{2\,N_{bck}}}
\end{eqnarray}
where $N_{\gamma}$ is the number of detected photons from DM
annihilations, and $N_{bck}$ the number of background photons.

For each annihilation of neutralino pairs, the differential
continuum photon spectrum  $dN_{\gamma}/dE_{\gamma}$ expected from
SUSY signals can be approximated~\cite{tasitsiomi,ams}:
\begin{equation}
\begin{split}
\frac{dN_{\gamma}}{dE_{\gamma}}\,=\,\frac{1}{m_{\chi}}\,
{\bigg\lgroup}\frac{10}{3}+\frac{5}{12}\left(\frac{E_{\gamma}}{m_{\chi}}\right)^{-3/2}
-\frac{5}{4}\left(\frac{E_{\gamma}}{m_{\chi}}\right)^{1/2}\\
-\frac{5}{2}\left(\frac{E_{\gamma}}{m_{\chi}}\right)^{-1/2}{\bigg\rgroup}
\end{split}
\label{eqn1}
\end{equation}
This leading-log approximation is obtained from a quark
fragmentation model for the total hadron spectrum using the Hill
spectrum. For more details, see~\cite{tasitsiomi}. On the other
hand, the number of continuum photons above the energy threshold
$E_{th}$ collected by the telescope array can be expressed in
terms of:
\begin{eqnarray}
N_{\gamma}(E_{\gamma} >
E_{th})\,=\,T_{obs}\,\int_{E_{th}}^{\infty}A_{eff}(E_{\gamma})\frac{d\Phi_{\gamma}}{dE_{\gamma}}dE_{\gamma}
\label{eqn2}
\end{eqnarray}
where $A_{eff}(E_{\gamma})$ corresponds to the effective area of
the H.E.S.S. instrument for given energy, zenith angle, etc.,
$T_{obs}$ the observation time, and $d\Phi_{\gamma}/dE_{\gamma}$
the differential $\gamma$-ray flux from DM particle annihilations
calculated with the formula given in Eq.~(\ref{eqn:flux}). Using
Eq.~(\ref{eqn1}) and Eq.~(\ref{eqn2}) and inserting the expression
of $d\Phi_{\gamma}/dE_{\gamma}$, after integration one can compute
the minimum detectable annihilation rate $\langle \sigma v
\rangle_{min}$ :
\begin{eqnarray}
\langle \sigma v
\rangle_{min}\,=\,\frac{4\pi}{\bar{J}(\Delta\Omega)\,\Delta\Omega}\frac{S\,m_{\chi}^2\,\sqrt{2\,N_{bck}}}
{\displaystyle\int_{E_{th}}^{\infty}
A_{eff}(E_{\gamma})\frac{dN_{\gamma}}{dE_{\gamma}}dE_{\gamma}}
\label{eqn4}
\end{eqnarray}
For the signal under consideration, the major sources of
background are protons, nuclei and electrons. Since the nuclei
background is subdominant compared to others, this background will
be neglected in what follows.

In the case of the hadronic background, the computation of the
number of background events to be detected is derived using the
following expression~\cite{bergstrom} :
\begin{eqnarray}
\frac{d\Phi_{had}}{d\Omega}(E > E_{th})\,=\,6.1\times
10^{-3}\epsilon_{had}\left(\frac{E_{th}}{1\,\rm GeV}\right)^{-1.7}
\nonumber\\
{\rm [cm^{-2}s^{-1}sr^{-1}]} \label{eqn5}
\end{eqnarray}
where $\epsilon_{had}$ represents the expected hadronic rejection
which is in the case of H.E.S.S. II of the order of 80\%. The
contribution from cosmic ray electrons
 that initiate showers, which is not distinguishable
from $\gamma$-rays, is~\cite{bergstrom} :
\begin{eqnarray}
\frac{d\Phi_{e^-}}{d\Omega}(E > E_{th})\,=\,3.0\times
10^{-2}\left(\frac{E_{th}}{1\,\rm GeV}\right)^{-2.3}
\nonumber\\
{\rm [cm^{-2}s^{-1}sr^{-1}]} \label{eqn6}
\end{eqnarray}

\section{Constraints from H.E.S.S. II on gamma-ray fluxes}
\label{sec:constraint} 
\begin{figure}[hb]
\mbox{\includegraphics[scale=0.45]{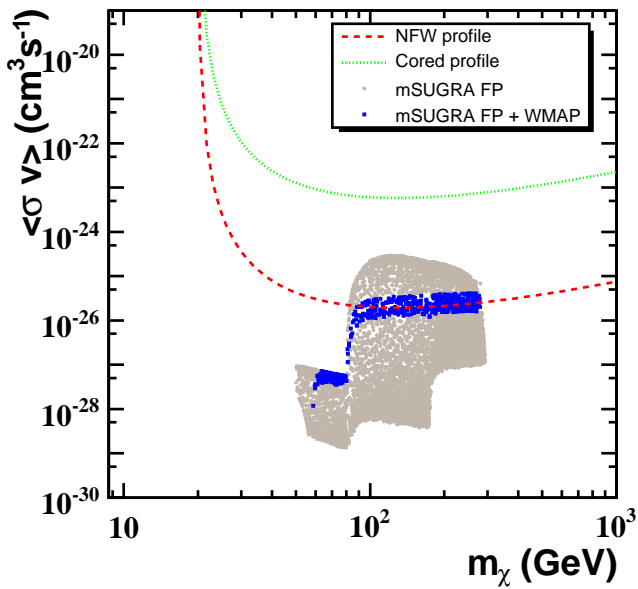}} \caption{5$\sigma$
exclusion limits for H.E.S.S. II on the velocity-weighted
cross-section $\langle \sigma v \rangle$ as a function of the
neutralino mass $m_{\chi}$ for the Galactic Center with a NFW
profile (dashed red line). Also indicated is the case of a cored
profile (dotted green line). The energy threshold is 20 GeV and
the observation time is 50 hours. mSUGRA models (grey points) from
the FP scanned region defined in Table~\ref{tab:scanFP} are
presented as well as those satisfying WMAP constraints on the CDM
relic density (blue points).}
 \label{fig:exclusionlimits}
\end{figure}

In the framework described in
Sec.~\ref{sec:susy}, the SUSY spectrum has been computed for
various configurations in the Focus Point region. The scanned
parameter region  is defined in Table~\ref{tab:scanFP}. For each
set of parameters, the integrated $\gamma$-ray flux has been
derived using  Eq.~(\ref{eqn:flux}) integrated above the energy
threshold. We assume as a benchmark halo shape a NFW profile which
corresponds to a standard distribution
\begin{table}[!ht]
\caption{\label{tab:scanFP}The case study parameter region in the
Focus point regime. The top mass $m_t$ is set to 175 GeV (see text
for discussions on sensitivity to $m_t$.)}
\begin{ruledtabular}
\begin{tabular}{ccc}
Parameter&Minimum& Maximum\\
\hline
\\
$m_0$ (GeV)&2200&5200\\
$m_{1/2}$ (GeV)&150&750\\
\multicolumn{3}{c}{\raisebox{0pt}[12pt][6pt]{$sign(\mu$)\,=\,+\,1}}\\
\multicolumn{3}{c}{\raisebox{0pt}[12pt][6pt]{$\tan\beta$ = 10}}\\
\multicolumn{3}{c}{\raisebox{0pt}[12pt][6pt]{$A_0$=0}}\\
\end{tabular}
\end{ruledtabular}
\end{table}
for dark matter. A second
less optimistic halo parametrization, presenting a core profile
with asymptotically flat velocity dispersion curve, has been also
considered. Table~\ref{tab:jbar} reports the value of the
astrophysical quantity
 $\bar{J}$ defined in Eq.~(\ref{eqn:jbar}) for the Galactic Center,
in the case of the two aforementioned halo profiles for
comparison. The resulting range of variation of $\bar{J}$ allows
us to estimate the magnitude of the astrophysical uncertainties
affecting the determination of the $\gamma$-ray flux from
neutralino annihilation. Fig.~\ref{fig:exclusionlimits}  presents
the 5$\sigma$ exclusion limit of H.E.S.S. II for the annihilation
cross-section using a given acceptance parametrization $A_{eff}
(E_{\gamma})$, calculated for a generic Galactic Center source
with a NFW and a cored DM profile respectively. Assuming an
observation time of 50 hours and a 20 GeV energy threshold, a
sensitivity as low as $\rm \mathcal{O}(10^{-26})\,cm^3s^{-1}$ for
the velocity-weighted annihilation
cross-section can be achieved in case of a NFW profile.
The SUSY models from Table~\ref{tab:scanFP} are also shown. In the
scanned region, neutralino masses range from 50 GeV up to 300 GeV
and velocity-weighted cross-sections from 10$^{-29}$ up to $\rm
\sim 10^{-25}~cm^3s^{-1}$.
\begin{table}[!hb]
\caption{\label{tab:jbar}Values of $\bar{J}(\Delta\Omega)$ for the
Galactic Center with a NFW and cored profiles
respectively~\cite{evans2} for the solid angle $\Delta \Omega = 2
\times 10^{-5}$ sr.}
\begin{ruledtabular}
\begin{tabular}{cccc}
&Profile&$\bar{J}$ ($\rm 10^{23} GeV^2cm^{-5}$)&\\
\hline
\\
&NFW &270&\\
&Core &0.7&\\
\\
\end{tabular}
\end{ruledtabular}
\end{table}
The lowest neutralino masses as well as annihilation
cross-sections are reached through the monoenergetic lines
resulting from loop-induced processes such as $\chi\chi
\rightarrow \gamma\gamma$, $\chi\chi \rightarrow \gamma Z$ and
$\chi\chi \rightarrow \gamma h$.
\begin{figure}[!ht]
\includegraphics[scale=0.45]{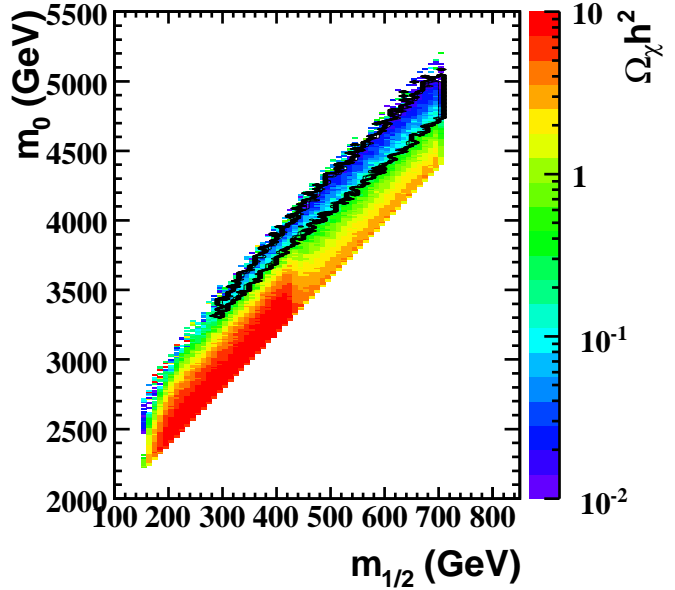}
\caption{Neutralino relic density, $\Omega_{\chi}h^2$, in the
(m$_0$,m$_{1/2}$) plane for the FP region defined in
Table~\ref{tab:scanFP} with A$_0$ = 0, tan$\beta$ = 10,
sign($\mu$) = +1 and m$_t$ = 175 GeV. Overlaid is the H.E.S.S. II
sensitivity for a 5$\sigma$ detection in 50 hours (black
contours). } \label{fig:fp10}
\end{figure}
A large fraction of the models obtained in the scan are within the
reach of H.E.S.S. II in the case of a NFW halo profile.
 This opens the possibility to test a
fraction of these models which satisfy constraints on the CDM
relic density coming from the Wilkinson Microwave Anisotropy Probe
(WMAP) \cite{wmap}. As for comparison, in the case of the cored
halo profile, the H.E.S.S. II sensitivity reaches $\rm \sim
10^{-23} cm^3s^{-1}$ and will not be able to constrain this
parameter space region.

\begin{figure*}
\begin{minipage}{175mm}
\mbox{\includegraphics[scale=0.45]{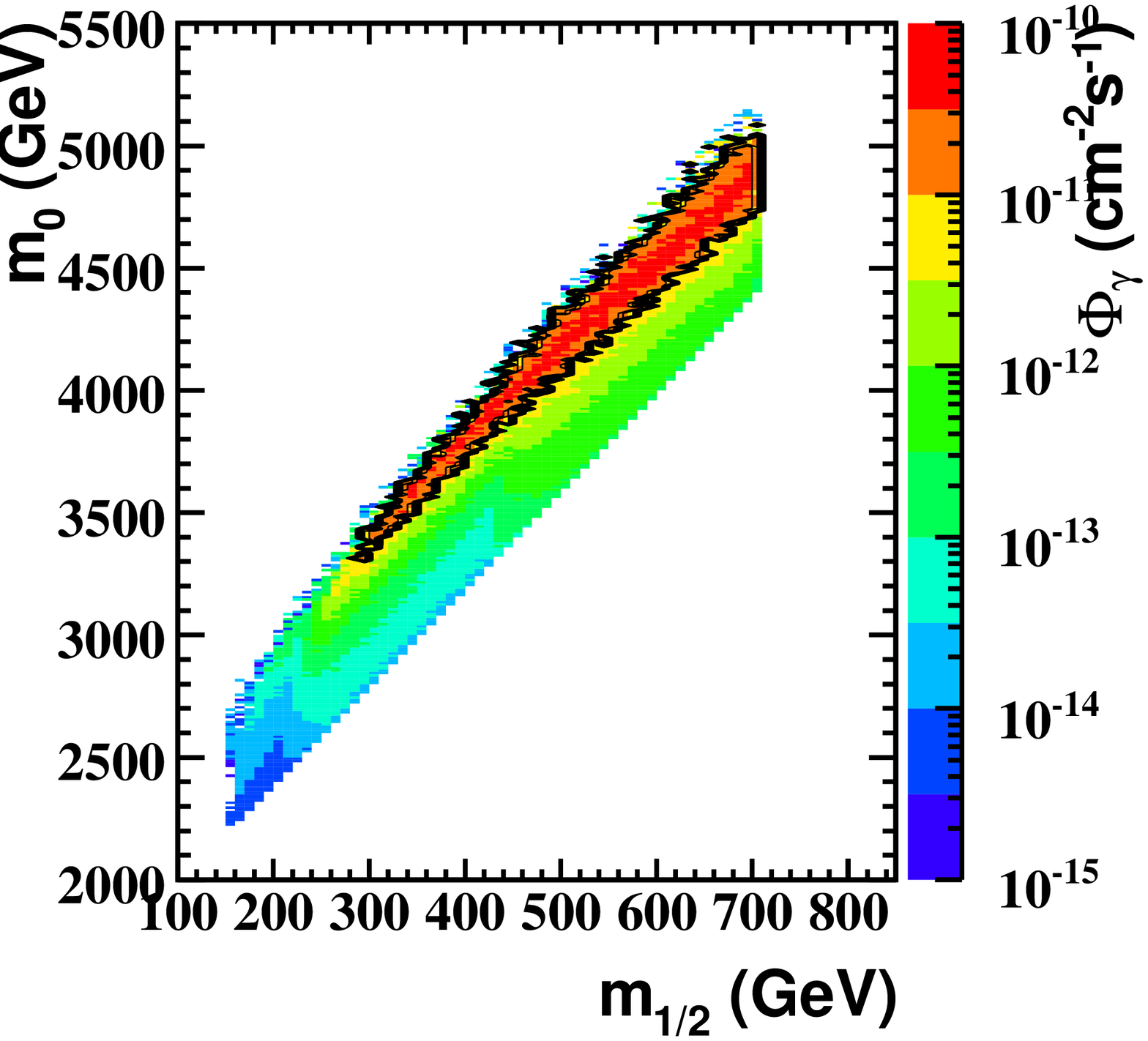}
\includegraphics[height=8.6cm,width=8.7cm]{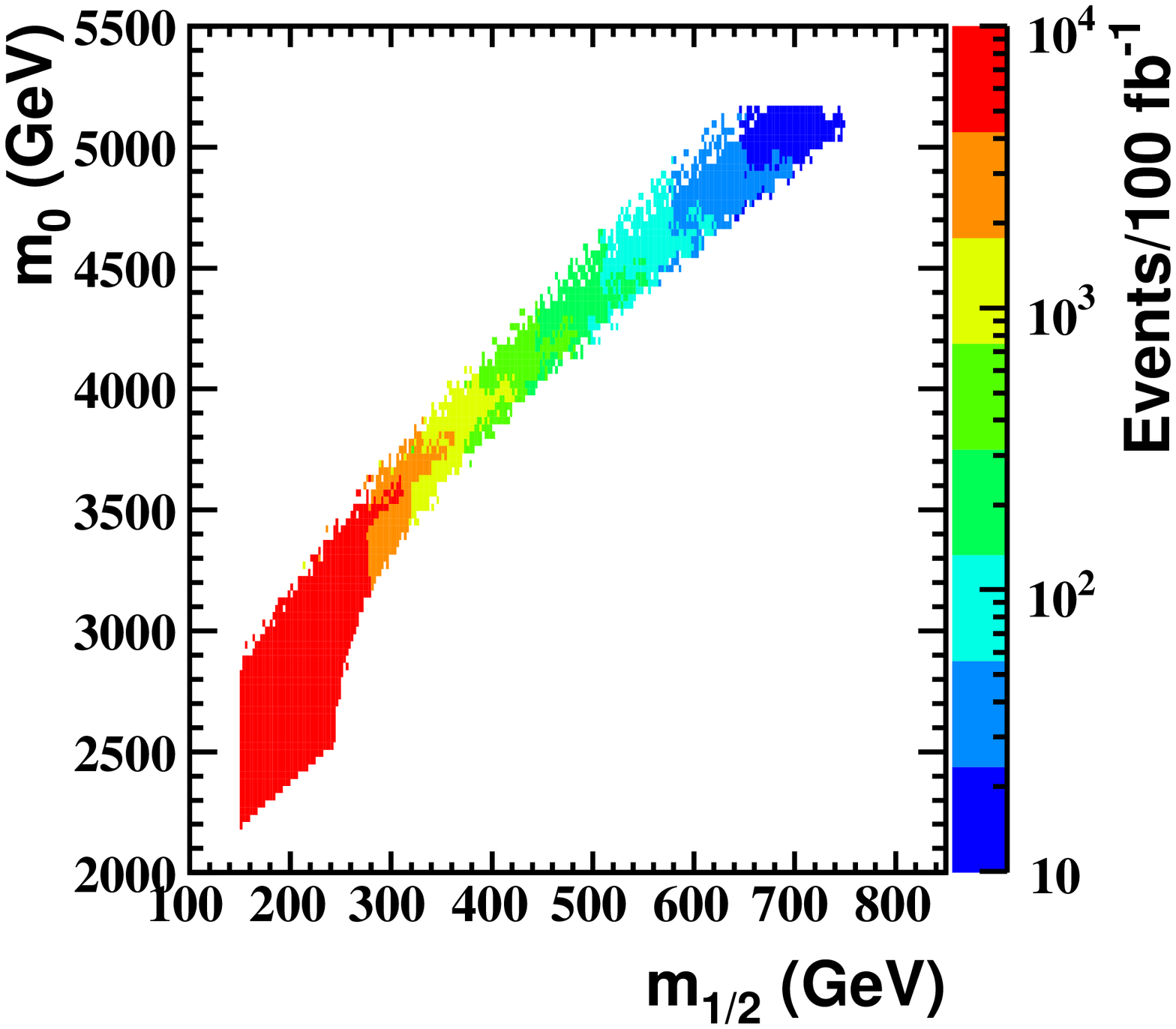}}
\caption{Left: Integrated $\gamma$-ray flux above 20 GeV,
$\Phi_{\gamma}$, in the (m$_0$,m$_{1/2}$) plane for the FP region
defined in Table~\ref{tab:scanFP} with $A_0$ = 0, $\tan\beta$ =
10, sign($\mu$) = +1 and $m_t$ = 175 GeV. Overlaid is the H.E.S.S.
II sensitivity for a 5$\sigma$ detection in 50 hours (black
contours). Right: Expected number of signal events before
selection efficiency cuts for the leptonic 3-body decay of
neutralinos (see Eq.~(\ref{eq:dec})) at the LHC for an integrated
luminosity of 100~fb$^{-1}$ in the $(m_0-m_{1/2})$ plane.
}\label{fig:m0mhalf}
\mbox{\includegraphics[scale=0.45]{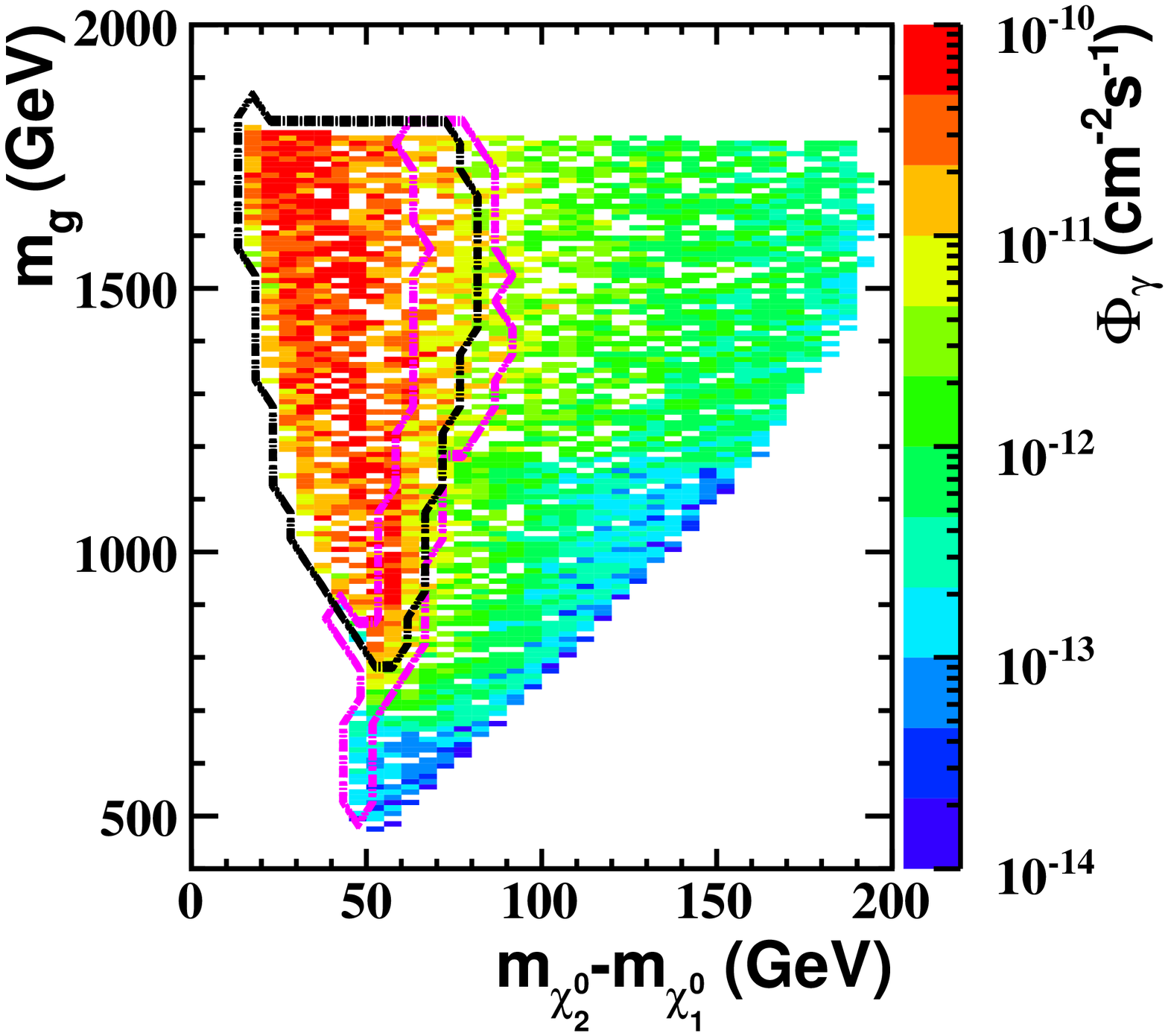}
\includegraphics[height=8.6cm,width=8.7cm]{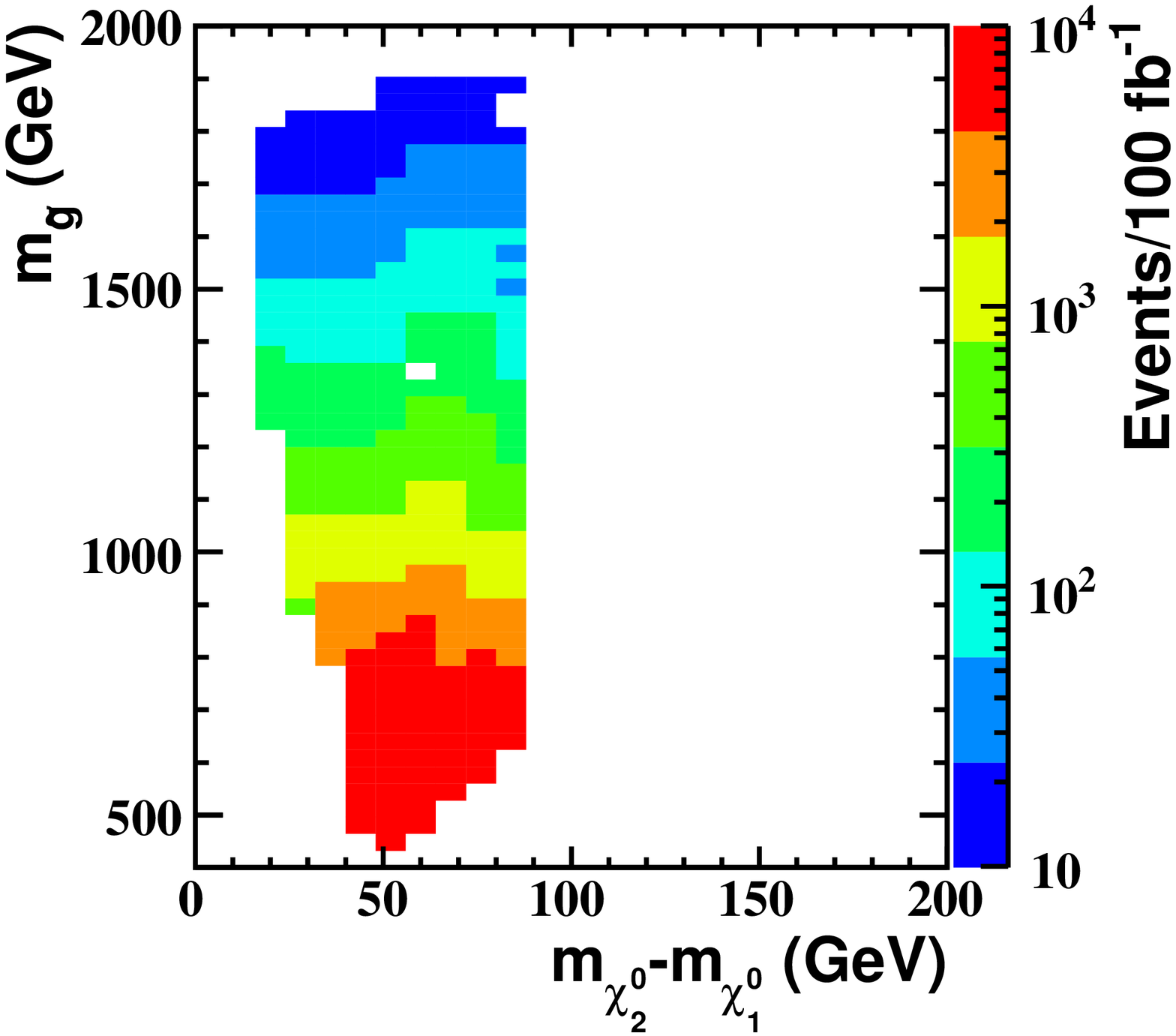}}
\caption{Left:Integrated $\gamma$-ray flux above 20 GeV in the FP
region defined in Table~\ref{tab:scanFP} in the plane of the
gluino mass, $m_{\tilde{g}}$, as a function of the mass difference
of the two lightest neutralinos $m_{\chi^0_2}-m_{\chi^0_1}$.
Overlaid are the H.E.S.S.-II sensitivity (solid black contour) and
the 5$\sigma$ allowed range from WMAP $\Omega_{CDM}h^2$
measurements (dashed purple contour). Right: Expected number of
signal events before selection efficiency cuts for the leptonic
3-body decay of neutralinos (see Eq.~(\ref{eq:dec})) at the LHC
for 100~fb$^{-1}$ in the ($m_{\chi^0_2}-m_{\chi^0_1}$,\,m$_{\tg}$)
plane.} \label{fig:gluinomass}
\end{minipage}
\end{figure*}

The same model set, defined in Table~\ref{tab:scanFP}, is shown in
the left-hand side panel of Fig.~\ref{fig:m0mhalf} in the
$(m_0-m_{1/2})$ plane for a NFW profile. The higgsino content of
the lightest neutralino $\chi$ implies large annihilation via
$W^{+}W^{-}$, $ZZ$ gauge bosons and substantial amount of
$\gamma$-rays in the final state. For illustration, a typical
value of m$_0$ as high as 3000 GeV is required to obtain a
satisfactory value of the neutralino relic density for $m_t$ = 175
GeV and $m_{1/2}$ = 300 GeV. Furthermore, as the annihilation
channels leading to $\gamma$-rays control also the thermal relic
density, we observe a strong (anti)correlation between the latter
and the $\gamma$-ray flux $\phi_{\gamma}$. As can be seen from
left panel of Fig.~\ref{fig:m0mhalf}, $\phi_{\gamma}$ spans 5
orders of magnitude, while  the relic density can vary rapidly
over 3 orders of magnitude in the considered parameter space
region as shown in Fig.~\ref{fig:fp10}.

Only a narrow region can accommodate the WMAP constraints on the
CDM relic density. In order to account for these constraints, we
highlight the models yielding a neutralino relic density lying in
the range $0.067 \leq \Omega_{\chi} h^2 \leq 0.156$ (corresponding
roughly to 5 standard deviations). The FP region is extremely
sensitive to the value of the top mass $\rm m_{t}$. Indeed, for
large $m_0$ values the steep running of the $m_{Hu}$ Higgs doublet
mass term translates into a high sensitivity of the $\mu$
parameter to $m_t$ through the requirement of radiative
electroweak symmetry breaking. As already mentioned in section 2,
one can then obtain sustantially different sparticle spectra when
using different RGE codes~\cite{Kraml}. It is well known that the
above-mentioned strong sensitivity can lead to dramatically
different $\Omega_{\chi} h^2$ values for a {\em fixed} choice of
the mSUGRA input parameters $m_0,m_{1/2}$, etc. Thus using
different codes can possibly require slight readjustment of the
input parameter values to retrieve the appropriate WMAP allowed
region. Nevertheless once a scan is performed over a reasonably
large domain in the $(m_0,m_{1/2})$ plane, the effect of using
different codes is simply to ``shift around" a bit the different
contours for the WMAP allowed ranges, eventually modifying
somewhat their shapes, which has been checked to some extent in
the present study using the code SuSpect~\cite{suspect}. Thus the
discrepancies in the sparticle spectrum in the focus region, as
obtained from the different publically available codes, should not
affect appreciably our main results, provided of course that the
same code is used consistently for the whole study.\footnote{More
precisely the most important differences occur in fact between
Isajet 7.69 on one side and
Suspect/SoftSuSy~\cite{softsusy}/Spheno~\cite{spheno} on the other
for the determination of the $\mu$ parameter, while the
differences among the latter three codes appear to be generally
much milder~\cite{Kraml}. It should be noted however that the
discrepancies in the LSP neutralino mass remain moderate even
though in the focus point region its higgsino component is not
negligible, since the neutralino $N_1$ component remains
essentially determined by the $U(1)_Y$ gaugino soft mass term
$M_1$ in this region, so that its mass is not very sensitive to
$\mu$.} However, we will occasionally present the output of two
versions of ISAJET, 7.69 and 7.71, in order to better illustrate
the range of reliability of the predicted observables.

The phenomenology of the FP region differs from the so-called bulk
and co-annihilation regions due to the large masses of the
sfermions. The scalar sector, including the CP even/odd and
charged Higgs,  lies
 in the few TeV range except for the lightest Higgs boson.
In performing the scanning of SUGRA parameters, we have checked
that a number of present experimental and theoretical constraints
are fulfilled. Apart from the direct lower limits on the sparticle
masses obtained at LEP or Tevatron~\cite{pdg} which are obviously
fulfilled in the range of parameters we consider here, there are
potentially some constraints from virtual supersymmmetric
contributions to low energy observables. The leading $\tilde
\chi^\pm \tilde t$ and $t H^\pm$ loop corrections to the branching
ratio for radiative $b$ decays, which can give in principle
interesting constraints on supersymmetric models, are completely
negligible in the present case due to the decoupling effects of
the very heavy sfermion and charged Higgs masses. Similarly,
supersymmetric contributions~\cite{amususy} to $a_\mu \equiv
(g_\mu-2)/2$, which would be typically enhanced for large values
of the combination $\mu \times \tan\beta$ through
charginos/sneutrinos (and more moderately neutralinos/smuons)
one-loop corrections, are suppressed in our case due to the very
large sfermion masses and moderate values of $\tan\beta$. In
particular this focus point region would be consistent with the
comparison between the measurements~\cite{gmuexp} and the standard
model theoretical predictions~\cite{gmuth} for $a_\mu$, if the
hadronic $\tau$-decay data (rather than the $e^+e^-$ annihilation
into hadrons) are used in the determination of the leading
hadronic contribution ~\cite{gmudis}.

The left panel of Fig.~\ref{fig:gluinomass} shows the range of
gluino masses, $m_{\tilde{g}}$, as a function of the two lightest
neutralino mass difference, $ m_{\chi^0_2}-m_{\chi^0_1}$, which is
a key observable for ATLAS (see section 5),  and the corresponding
$\gamma$-ray flux. The latter observable is indirectly sensitive
to the neutralino mass difference and somewhat in a model
dependent way. In our mSUGRA constrained scenario, and given the
LSP mass range and large $m_0$ values considered here, the $ZZ$
and $W^+W^-$ annihilation channels are open and become actually
dominant as compared to the fermion pair ($f\bar{f}$) final state
channels. This is a combined effect of, on one hand the
suppression of $f\bar{f}$ final states  due to $t-$channel
exchange of very heavy sfermions, and on the other hand a
substantial higgsino component of the LSP enhancing its coupling
to Z's and W's. A larger $m_{\chi^0_2}-m_{\chi^0_1}$  corresponds
to  heavier neutralinos (resp. charginos) which are exchanged in
the $t-$channel, and thus reduces the $ZZ$  (resp. $W^+W^-$) final
states, implying an overall reduction of the $\gamma$-ray fluxes.
The sensitivity of H.E.S.S. II is displayed as solid black
contour. Models satisfying WMAP constraints on $\Omega_{\chi}h^2$
are also highlighted as dashed purple contours. Gluino masses
lying in the range 800-1800 GeV are within the reach of H.E.S.S.
II sensitivity for neutralino mass differences up to 80 GeV. Given
the WMAP constraints, a significant fraction of cosmologically
interesting SUSY models will be tested with H.E.S.S. II.

\section{LHC measurements}
\label{sec:atlas} The scan described in the previous sections was
performed over a region of mSUGRA space for which $m_{\tilde
g}\le$ 1800~GeV. Various studies are available in the
literature~\cite{atltdr,cmstdr2,Tovey:2002jc,Baer:2003wx} which
demonstrate that for an integrated luminosity of $\sim
100$~fb$^{-1}$ inclusive searches for events including multiple
jets and/or leptons and missing transverse energy would allow the
discovery at the LHC of gluino production up to masses of $\sim
1800$~GeV, thus covering most of the investigated region.\par
Beyond the mere discovery of a signal, the LHC should be able to
perform measurements of the SUSY spectrum which would allow the
experiments to put constraints on the predicted relic
density~\cite{Polesello:2004qy,Nojiri:2005ph,Baltz:2006fm}. One
point in the region addressed here was studied in detail
in~\cite{desanctis}, and the possible ways of extracting a
parameter measurement were investigated. It turned out that the
cleanest signal for parameter measurements would be the study of
the three-body decay
\begin{eqnarray}
 \chi^0_{2(3)}\rightarrow\ell^+\ell^-\chi^0_1
\label{eq:dec}
\end{eqnarray}
where the $\chi^0_{2(3)}$ would be produced in the decay
$\tg\rightarrow qq\chi^0_{2(3)}$. Alternatively the same signal
could be searched for through the direct production of neutralinos
in association with charginos in proton-proton interactions.

This chain would give a clean final state signature~\cite{atltdr}
with two opposite-sign same-flavour leptons. The kinematics of the
three-body decay imply that the invariant mass of any two
particles in the decay must be smaller than the difference between
the mass of the mother particle and the mass of the remaining
particle. Therefore, from the observation of one (or more)
kinematic  end-points in the invariant mass distribution of the
two leptons from the decay given in Eq.~(\ref{eq:dec}) the mass
difference $m_{\chi^0_i}-m_{\chi^0_1}$ can be measured. The error
on this measurement has a systematic component, arising from
lepton selection cuts, and a statistical component which will
scale both with the value of the mass difference and with the
number of events observed. This latter component will vary over
the SUSY parameter space due to variations in the mass spectrum
and in the total SUSY production cross-section. For the model
studied in~\cite{desanctis}, a precision of order 1\% is
obtained, which reflects the excellent measurement capabilities of
the LHC experiments for leptons.

We have investigated the LHC reach for this measurement for the
region of mSUGRA space defined in section 4, by calculating the
number of events expected at the LHC for the decay given in
Eq.~(\ref{eq:dec}) for an integrated luminosity of 100~fb$^{-1}$.
The number of events shown are independent of the detector
assumptions. For the quoted selection efficiencies we refer to an
ATLAS analysis in parametrised simulation. The results are valid
for the generic performance of an LHC detector and are expected to
be equivalent for CMS. We used ISAJET 7.71~\cite{Paige:2003mg} for
the calculation of spectra and branching fractions and
Prospino~\cite{Beenakker:1996ch} for the NLO cross-section for
gluino production. The results are shown in the right-hand side
panel of Fig.~\ref{fig:m0mhalf} in the $(m_0-m_{1/2})$ plane
before selection efficiency cuts. The sharp cut on the right-hand
side is due to the fact that we do not consider models for which
$m_{\chi^0_2}-m_{\chi^0_1}>85$~GeV, at which point the kinematic
end point in the two-lepton invariant mass distribution starts to
be masked by the Z peak that arises from the decay in two leptons.
When the decay $\chi^0_{2}\rightarrow Z\chi^0_1$ becomes
kinematically accessible (i.e. when the neutralino mass difference
exceeds the Z mass), the previous kinematic signature resulting
from a three-body decay is lost, and the two body process that
occurs instead does not allow us to make an unambiguous and model
independent measurement of the neutralino mass difference. This
two body process will dominate up to the point where a neutralino
decay channel featuring a light higgs h, $\chi^0_{2}\rightarrow
h\chi^0_1$, becomes open. The higgs decays mostly into $\rm
b\bar{b}$ pairs, and the extraction of the $h\rightarrow b
\bar{b}$ peak above the large combinatorial of $b$ jets from
gluino decays is very challenging. For the three-body decay of the
neutralinos to leptons, in the analysis of~\cite{desanctis} an
efficiency of $\sim$ 25\% is achieved for the signal, the SUSY
background is $\sim$ 0.6 times the signal, and a SM background of
$\sim$ 250 events is expected for 100~fb$^{-1}$. The SM background
is dominated by $\bar tt$ production. This background, as well as
the SUSY backgrounds, can be easily evaluated from the data
themselves. In fact the two leptons from the three-body decays of
the neutralino, being from the decay of a virtual Z, must have the
same flavour and opposite sign. The two isolated leptons from
$\bar tt$ are produced from the decays of two W and can therefore
have any flavour combination. It is therefore easy to
statistically subtract the background by subtracting the invariant
mass distribution of $e^\pm\mu^\mp$ pairs from the sum of the
distributions for $e^\pm e^\mp$ and $\mu^\pm \mu^\mp$ pairs.
Equivalent considerations are valid for the SUSY backgrounds,
where the flavours of the two leptons are uncorrelated. Assuming
that the efficiency of the analysis cuts and ratio between signal
and SUSY background do not vary dramatically in the investigated
region, a signal of $\sim$350 events would be needed for the
observation at 5$\sigma$ level for an integrated luminosity of
100~fb$^{-1}$. The number of signal events is essentially
determined by the gluino mass, as is clear in the right panel of
Fig.~\ref{fig:gluinomass}, where the number of signal events is
given on the ($m_{\chi^0_2}-m_{\chi^0_1},m_{\tg}$) plane. The
opposite sign - same flavour two-lepton signal should therefore be
observable for $m_{\tg}\le 1300$~GeV.

\begin{table*}[!ht]
\caption{\label{tab:su2} Values of the neutralino relic density
$\Omega_{\chi}\,h^2$, the $\gamma$-ray flux $\Phi_{\gamma}$, the
number of events per 100 fb$^{-1}$ after selection efficiency cuts
for NLSP 3-body leptonic decays, the mass difference of the two
lightest neutralinos and the gluino mass, for the SUSY point
defined in Eq.~(\ref{eqn:point}). The two quoted values for each
observable correspond respectively to the output of ISAJET 7.69
and 7.71 (see text for details).}
\begin{ruledtabular}
\begin{tabular}{cc|cc|cc|cc|c}
\multicolumn{2}{l|}{Relic density $\Omega_{\chi}h^2$}&\multicolumn{2}{l|}{$\gamma$-ray flux $\Phi_{\gamma}$ (cm$^{-2}$s$^{-1}$)}&\multicolumn{2}{l|}{Events per 100 fb$^{-1}$}& \multicolumn{2}{l|}{$m_{\chi_2^0}-m_{\chi_1^0}$ (GeV)} & $m_{\tilde{g}}$ (GeV)\\
\hline
&&&&&&&&\\
0.06&0.16&6.6$\times$10$^{-12}$&2.3$\times$10$^{-12}$&855&468&55&88&852\\
&&&&&&&&\\
\end{tabular}
\end{ruledtabular}
\end{table*}
\section{Discussion of the complementarity}
\label{sec:discussion} We discuss here the complementarity of the
indirect detection with the phase II of H.E.S.S. and the ATLAS
searches at LHC in the focus point region.

The potentialities of the two experiments in the FP region can be
readily compared through the integrated $\gamma$-ray flux and the
expected number of signal events for Next-to-LSP 3-body leptonic
decays, in the $(m_0,m_{1/2})$ plane, displayed respectively in
the left and right panels of Fig.~\ref{fig:m0mhalf}. The left-hand
side plot of Fig.~\ref{fig:m0mhalf} shows the H.E.S.S.
$\gamma$-ray flux contours for a 5$\sigma$ detection. The H.E.S.S.
$\gamma$-ray integrated flux sensitivity which is at the level of
$\rm \sim 10^{-12}\,cm^{-2}s^{-1}$, allows us to test SUSY models
characterized by universal scalar soft mass parameters above
 3300 GeV and gaugino soft mass parameters above 280 GeV. From the number of
signal events for 100 fb$^{-1}$ in the 3-body leptonic decay shown
in the right panel of Fig.~\ref{fig:m0mhalf}, the ATLAS
sensitivity will allow us to investigate $m_0$ lower than 4200 GeV
and $m_{1/2}$ as high as 550 GeV. Thus the combination of the
sensitivity of each experiment allows us to cover the overall
region. This point highlights the complementarity between the
indirect detection of dark matter and the SUSY searches at LHC.
Moreover, an overlap is obtained in a small region characterized
by scalar soft masses in the 3300-4200 GeV range and gaugino soft
masses between 280 and 550 GeV. In case of a signal discovery
consistent with this region, complementary cross-checks between
the two experiments would lead to further constraints on the
allowed parameter space in the $(m_0,m_{1/2})$ plane.

For further illustration, we have chosen in
Fig.~\ref{fig:gluinomass} a given point in the parameter space:
\begin{eqnarray} \label{eqn:point}
 m_0 = 3350\,{\rm GeV}, m_{1/2} = 300\,{\rm GeV}, A_0 =
0\,{\rm GeV}, \nonumber\\
\mu > 0, \tan \beta = 10
\end{eqnarray}
\noindent This point, with the top mass set to 175 GeV and the
mass spectrum computed with ISAJET, is close to the point
(sometimes dubbed SU(2) ) which has been chosen by the ATLAS
experiment for a detailed study~\cite{desanctis}.  For each
quantity reported in Table~\ref{tab:su2}, two values are given
corresponding to the output spectra of ISAJET7.69 and ISAJET7.71.
This allows to give an estimate for the theoretical uncertainties.
As can be seen from Fig.~\ref{fig:gluinomass} and
Table~\ref{tab:su2}, this particular point, which is within the
reach of ATLAS, is also within the reach of H.E.S.S. II provided a
typical 50 hours observation time of a Galactic Center type
source.

As explained in section~\ref{sec:atlas}, a key parameter in the
ATLAS sensitivity is the gluino mass. The sensitivity curves of
each experiment allow to constrain different regions in the plane
of the gluino mass versus the mass difference of the two lightest
neutralinos as shown in Fig.~\ref{fig:gluinomass}.
As can be seen,
 one anticipates
lower flux values for larger neutralino mass differences. Given an
astrophysical model for the source (see section 4), H.E.S.S. II
would thus put lower limits on the neutralino mass differences
which can be further refined with some knowledge of the gluino
mass. Moreover, given the sensitivity zone of H.E.S.S. II, one
finds that a larger  gluino mass would allow the exploration of a
larger fraction of the parameter space. In contrast, and as
illustrated in from Fig.~\ref{fig:gluinomass}, ATLAS will have
larger sensitivities for lighter gluinos. In any case, and
irrespective of the gluino mass, the extraction of information
from  ATLAS measurements becomes problematic for a neutralino mass
difference exceeding 85 GeV. Finally, comparing the two considered
observables in the panels of Fig.~\ref{fig:gluinomass}
one notes a reversed sensitivity to the mass
parameters: ATLAS would provide a strong constraint on the gluino
mass whereas the indirect detection is more sensitive to the
neutralino mass differences. Orthogonal constraints can thus be
obtained  by indirect detection and collider searches, and their
possible combination would reduce significantly the allowed
parameter space.

\section{Conclusion and perspectives}
In this paper we have examined the complementarity of two
approaches to unravel signatures of dark matter particles in the
case where the latter are neutralino LSPs in the Focus Point
region of the mSUGRA scenario. Searches at colliders in this
region which is difficult to investigate as compared to the
so-called bulk region, will benefit from complementary input from
indirect searches through $\gamma$-rays\footnote{We note that
searches in the bulk region are easier due to the fact that the
lower value of $m_0$ ensures that squarks and sleptons will be
visible at the LHC. The greater range of visible sparticles will
provide a greater knowledge of the sparticle mass spectrum.}.  In
the present study we used the most realistic simulations for both
ATLAS and H.E.S.S. II for which we considered the Galactic Center
as a benchmark astrophysical DM source. We found that the two
experiments will typically explore two different regions of the
studied parameter space with some possible overlap, allowing in
principle complementary analyses. We illustrated these features in
terms of the universal soft parameters as well as in terms of the
physical masses, the first corresponding to a top-down
model-dependent approach while the second illustrates the more
challenging model-independent one. For instance, in the scanned
range of neutralino LSP masses from 50 to 300 GeV, ATLAS will be
sensitive to gluinos lighter than 1300 GeV whereas H.E.S.S. II,
with a flux sensitivity of the order of $\rm 10^{-12}
cm^{-2}s^{-1}$, will be able to cover all the region above 1 TeV.
Furthermore, the mass difference of the two lightest neutralinos
which is a key observable for ATLAS, is interestingly found to be
a sensitive parameter for the $\gamma$-ray fluxes that H.E.S.S. II
can observe or constrain.

More generally, in the very near future, H.E.S.S. II and LHC will
be two major experiments allowing us to probe the supersymmetric
hypothesis for dark matter with an, up to now, unequalled
capability. Neutralino masses in the yet uncovered 100 GeV range
will be accessible to H.E.S.S. II thus filling the energy gap
between the current H.E.S.S. I and the GLAST satellite experiment
to be launched in Spring 2008. Finally, if SUSY signals are
discovered at the LHC, the cross-sections, branching ratios and
masses of the new particles determined with a few fb$^{-1}$
luminosity, would provide valuable search windows both for direct
and indirect supersymmetric dark matter detection. Of course,
substantial systematic uncertainties arise on the astrophysical
parameters in indirect searches, which can be reduced through an
improved knowledge of the halo profiles of the observed sources.
Direct detection also suffers from systematic effects related to
astrophysical and nuclear parameter uncertainties which affect the
constraints on the nucleon-WIMP cross-sections. The more
model-independent measurements of the SUSY parameters will require
data samples collected with hundreds of fb$^{-1}$ luminosity at
the LHC, as well as independent input from indirect (or direct)
dark matter searches as illustrated in this paper.


\begin{thebibliography}{99}
\bibitem{goldberg} H. Goldberg, Phys. Rev. Lett. 50 (1983) 1419;
J. Ellis {\it al.}, \NPB{238}{1984}{453}.
\bibitem{jungman} for reviews, see for instance
G. Jungman, M. Kamionkowski and K. Griest, \PR{267}{1996}{195},
and G. Bertone, D. Hooper and J. Silk, \PR{405}{2005}{279}.
\bibitem{hess} http://www.mpi-hd.mpg.de/hfm/HESS/HESS.html
\bibitem{magic} http://hegra1.mppmu.mpg.de/MAGICWeb/
\bibitem{veritas} http://veritas.sao.arizona.edu/index.html
\bibitem{ams0} http://ams.cern.ch/
\bibitem{glast} http://www-glast-stanford.edu/
\bibitem{pamela} http://wizard.roma2.infn.it/pamela/
\bibitem{weiglein} LHC/LC Study Group: G. Weiglein et al,
\PR{426}{2006}{47}. 
\bibitem{Baltz:2006fm}
  E.~A.~Baltz, M.~Battaglia, M.~E.~Peskin and T.~Wizansky,
  Phys.\ Rev.\  D {\bf 74} (2006) 103521.
\bibitem{Hooper:2006wv}
  D.~Hooper and A.~M.~Taylor,
  JCAP {\bf 0703} (2007) 017.
\bibitem{Barger:2007xf}
  V.~Barger, W.~Y.~Keung, G.~Shaughnessy and A.~Tregre,
  Phys.\ Rev.\  D {\bf 76} (2007) 095008.
\bibitem{FMM} J. Feng, K. Matchev and T. Moroi, Phys.\ Rev.\  Lett. {\bf 84} (2000) 2322 and
Phys.\ Rev.\  D {\bf 61} (2000) 075005.
\bibitem{FMW} J. Feng, K. Matchev and F. Wilczek Phys.\  Lett.\ B {\bf 482} (2000) 388 and
Phys.\ Rev.\  D {\bf 63} (2001) 045024.
\bibitem{BKPU} H. Baer {\it et al}, JHEP {\bf 0510} (2005) 020 and references
therein. 
\bibitem{BKTplus} H. Baer, T. Krupovnickas and X. Tata, JHEP
{\bf 0307} (2003) 020; B.~C.~Allanach, S.~Kraml and W.~Porod, JHEP
{\bf 0303} (2003) 016; G.~Belanger, S.~Kraml and A.~Pukhov,
\PRD{72}{2005}{015003}.
\bibitem{isajet} Isajet v7.72, H. Baer {\it et al.}, arXiv:hep-ph/0312045.
\bibitem{suspect} SuSpect 2.34, A. Djouadi, J.-L. Kneur and G. Moultaka,
Comput. Phys. Commun. {\bf 176} (2007) 426.
\bibitem{darksusy} P. Gondolo {\it et al.}, JCAP {\bf 0407} (2004) 008.
\bibitem{bergstrom} L. Bergstr\"om, P. Ullio and J. Buckley, \APP{9}{1998}{137}.
\bibitem{baltz} E.~A.~Baltz, C.~Briot, P.~Salati, R.~Taillet and J.~Silk, Phys.\ Rev.\  D {\bf 61} (1999) 023514.
\bibitem{fornengo} N. Fornengo {\it et al.}, \PRD{70}{2004}{103529}.
\bibitem{tasitsiomi0} A. Tasitsiomi {\it et al.}, \APP{21}{2004}{637}.
\bibitem{evans1} N. W. Evans, \MNRAS{260}{1998}{190}.
\bibitem{nfw} J. Navarro, C. Frenk and S. White, \APJ{462}{1996}{563}.
\bibitem{moore} B. Moore {\it et al.}, Phys.\ Rev.\  D {\bf 64} (2001) 063508.
\bibitem{binney1} J. J. Binney and N. W. Evans,
\MNRAS{327}{2001}{L27}.
\bibitem{debattista} V. Debattista and J. A. Selwood, \APJ{493}{1998}{L5}.
\bibitem{binney2} J. J. Binney, O. E. Gerhard and J. Silk,
\MNRAS{321}{2001}{471}.
\bibitem{mcmillan} P. McMillan and W. Dehnen,
\MNRAS{363}{2005}{1205}.
\bibitem{aharonian1} F. A. Aharonian, {\it et al.} (H.E.S.S. Collaboration),
\AA{425}{2004}{L13}. 
\bibitem{aharonian3} F. A. Aharonian {\it et al.} (H.E.S.S. Collaboration),
\NATURE{439}{2006}{695}. 
\bibitem{evans2} N. W. Evans, F. Ferrer and S. Sarkar, Phys.\ Rev.\  D {\bf 69} (2004) 123501. 
\bibitem{sgrdwarf} F. A. Aharonian {\it et al.} (H.E.S.S. Collaboration),
\APP{29}{2008}{55}.
\bibitem{crabe} F. A. Aharonian {\it et al.} (H.E.S.S.
collaboration), \AA{457}{2006}{899}.
\bibitem{punch} M. Punch (for the H.E.S.S. collaboration), Proc.
of Cherenkov 2005, Towards a Network of Atmospheric Cherenkov
Detectors VI, April 2005, Ecole Polytechnique, Palaiseau, France.
\bibitem{lima}  T. P. Li and Y. Q. Ma, \APJ{272}{1983}{317}.
\bibitem{tasitsiomi}  A. Tasitsiomi and A. V. Olinto,
\PRD{66}{2002}{083006}. 
\bibitem{wmap} http://map.gsfc.nasa.gov/
\bibitem{ams} A. Jacholkowska {\it et al.}, Phys.\ Rev.\  D {\bf 74} (2006) 023518. 
\bibitem{Kraml} http://cern.ch/kraml/comparison/ and the third reference
in \cite{BKTplus}.
\bibitem{softsusy} B. C. Allanach, Comput. Phys. Commun. 143 (2002) 305.
\bibitem{spheno} W. Porod, Comput. Phys. Commun.  153 (2003) 275.
\bibitem{pdg} Particle Data Group, J. of Phys. G 33 (2006) 1.
\bibitem{amususy} S. P. Martin and J. D. Wells,
\PRD{64}{2001}{035003};
G. Degrassi and G. F. Giudice,
\PRD{58}{1998}{053007}.
\bibitem{gmuexp} Muon $g-2$ Collaboration, G.W. Bennett et al., \PRL{89}{2002}{101804}, Erratum--ibid. \PRL{89}{2002}{129903}
and Phys. Rev. Lett. {\bf 92} (2004) 161802.
\bibitem{gmuth} M. Davier {\it et al.},
Eur. Phys. J. {\bf C31}, 503 (2003); 
K.~Hagiwara, A.~D.~Martin, D.~Nomura and T.~Teubner,
\PRD{69}{2004}{093003}.
\bibitem{gmudis} See e.g. for a recent review M. Passera, talk given at
the Tau06 Workshop, Pisa, Sept. 2006, hep-ph/0702027.
\bibitem{desanctis}
U. De Sanctis {\it al.},
Eur.\ Phys.\ J.\  C {\bf 52} (2007) 743.
\bibitem{atltdr} ATLAS Collaboration, {\em ATLAS detector and
physics\\
performance Technical Design Report}, CERN/LHCC 99-14/15 (1999);\\
http://atlas.web.cern.ch/Atlas/GROUPS/PHYSICS/\\TDR/access.html.
\bibitem{cmstdr2}
CMS Collaboration {\em CMS physics : Technical Design Report v.2 :
Physics performance} CERN-LHCC-2006-021 (2006);
http://cdsweb.cern.ch/search.py?recid=942733.
\bibitem{Tovey:2002jc}
  D.~R.~Tovey,
  Eur.\ Phys.\ J.\ C {\bf 4}  (2002) N4.
\bibitem{Baer:2003wx}
  H.~Baer {\it et al.},
  JHEP {\bf 0306}  (2003) 054.
\bibitem{Polesello:2004qy}
  G.~Polesello and D.~R.~Tovey,
  JHEP {\bf 0405}  (2004) 071.
\bibitem{Nojiri:2005ph}
  M.~M.~Nojiri, G.~Polesello and D.~R.~Tovey,
  JHEP {\bf 0603}  (2006) 063.
\bibitem{Paige:2003mg}
  H.~Baer {\it et al.},
  arXiv:hep-ph/0312045.
\bibitem{Beenakker:1996ch}
  W.~Beenakker {\it et al.},
  Nucl.\ Phys.\  B {\bf 492} (1997) 51.
\end{thebibliography}
\end{document}